# Exploiting Spectral Leakage for Spectrogram Frequency Super-resolution

Ray Maleh, Frank A. Boyle *Member, IEEE*

***Abstract*— The spectrogram is a classical DSP tool used to view signals in both time and frequency. Unfortunately, the Heisenberg Uncertainty Principal limits our ability to use them for detecting and measuring narrowband signal modulation in wideband environments. On a spectrogram, instantaneous frequency can only be measured to the nearest bin without additional interpolation. This work presents a novel technique for extracting higher accuracy frequency estimates. Whereas most practitioners seek to suppress spectral leakage, we use mismatched windows to exploit such artifacts in order to produce super-resolved spectral displays. We present a derivation of our methodology and exhibit several interesting examples.**

**Index Terms— time-frequency analysis, Fourier Transform, spectrogram, spectral leakage, super-resolution**

## I. SPECTROGRAMS AND HEISENBERG'S UNCERTAINTY PRINCIPLE

Inspired by the Short-Time Fourier Transform, the spectrogram (or periodogram) is a signal processing tool that is often used to view signal contents simultaneously in both time and frequency. Given a discrete signal $x[n]$ of length $N_S$, a window $w[n]$ of length $N$, and the assumption that $N$ divides equally into $N_S$, we formally define the spectrogram (with 50% overlap) as:

$$X(\ell,k)=\sum_{n=0}^{N-1}x\left[n+\frac{\ell N}{2}\right]w[n]\exp\left(\frac{-j2\pi kn}{N}\right) \quad (1)$$

In essence, the spectrogram is a matrix whose columns consist of moving DFTs. A graphical depiction of the construction of a spectrogram is show below in Fig. 1. A key observation is that shorter DFT window sizes (increased time resolution) means that fewer DFT bins will be available (decreased frequency resolution). This is consistent with the Heisenberg Uncertainty Principle, which formally states that one cannot simultaneously have full time and frequency resolution.

Now suppose the "CW" in Fig. 1 contains some very narrow-band FM modulation that we wish to recover. The textbook approach to this problem is to tune and filter the signal and differentiate its unwrapped phase. Other approaches for estimating instantaneous frequency, including LMS/ML methods, zero-crossing techniques, quadratic interpolations, and others, are presented in [1],[2], and [3]. In Section II, we propose an interesting alternative methodology that exploits spectral leakage in order to visualize a signal and its super-resolved modulation on the same spectrogram surface. In Section 0, we present several examples involving simulated and real signals. Then, in Sections IV and V, we offer a detailed mathematical derivation of the results of this paper including a Cramér-Rao Bound analysis.

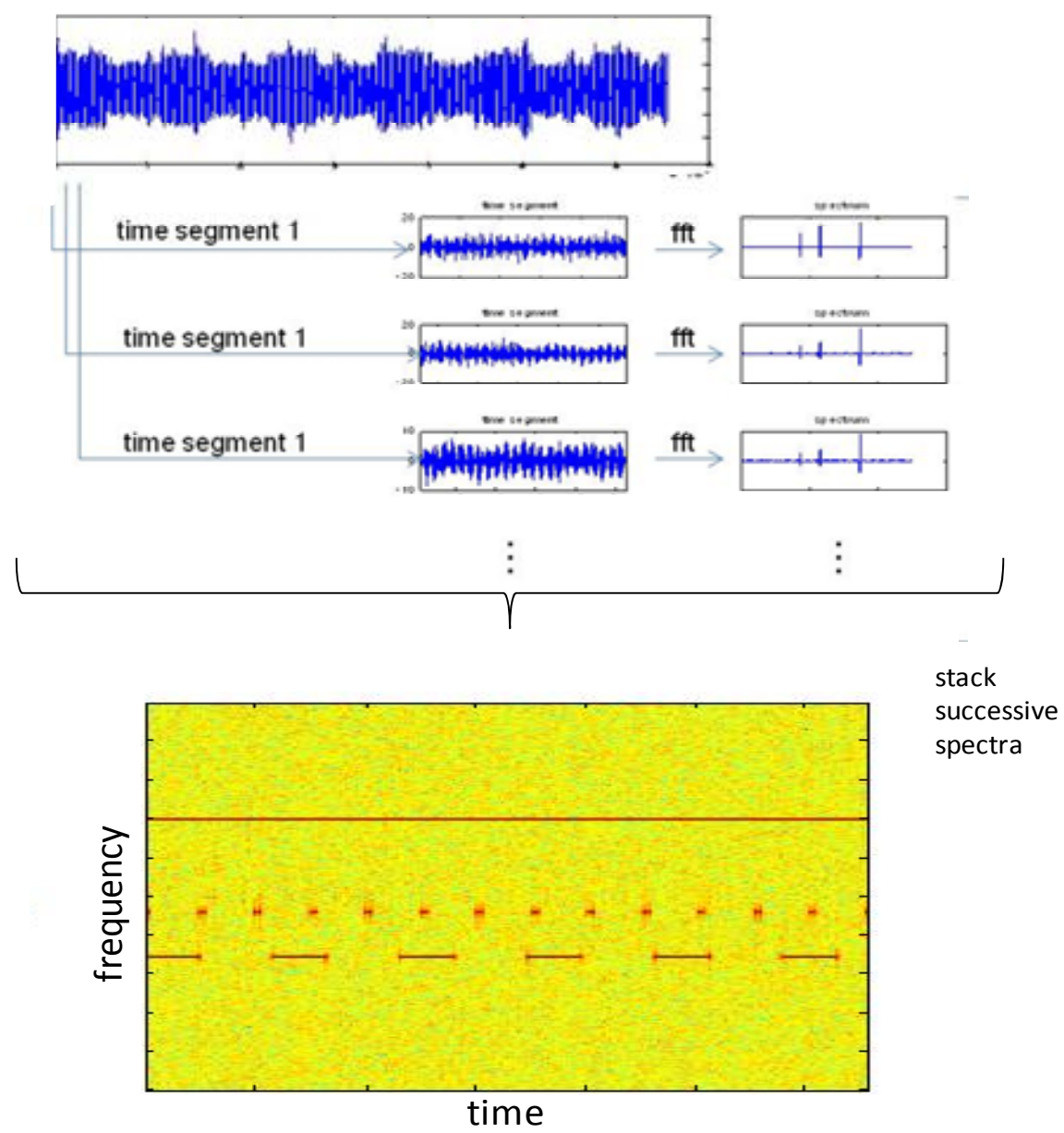

Fig. 1. Example of spectrogram showing a CW signal and two OOK signals.

## II. THE MISMATCHED TIME WINDOW SPECTROGRAM

It was observed *per accidens* that spectrograms of narrowband signals utilizing rectangular windows with missing (zeroed) entries contained nulls whose instantaneous frequencies matched the underlying signal modulation. An example of this is shown in Fig. 2 where two super-resolved replicas of the original narrow-band signal are present. In order to better characterize this phenomenon, we define the

The authors are with L-3 Communications Integrated Systems, Mission Integration Division, 10001 Jack Finney Blvd. Greenville, TX 75402, USA (e-mail: Ray.Maleh@L-3com.com, Frank.A.Boyle@L-3com.com,), Phone: 903-408-2605.

standard mismatched time window (MMTW) $w$[n] as shown below in (2):

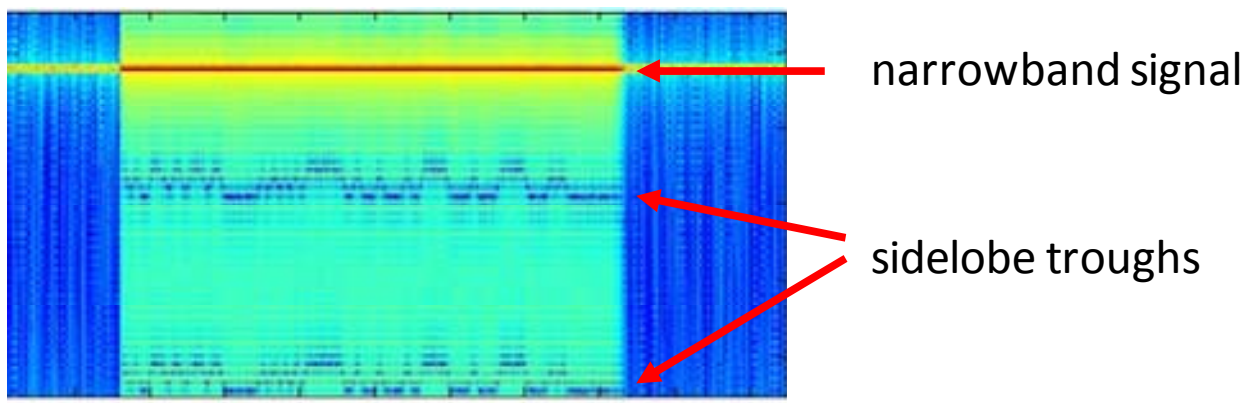

Fig. 2. Curious-looking spectrogram revealing the "microscopic" modulation of the original narrow-band signal.

$$w[n] = \begin{cases} 0 & \text{if } n = 0 \\ 1 & \text{otherwise} \end{cases} \tag{2}$$

We consider a short signal of length $N$ (i.e., one time-block of a spectrogram) which consists of a single tone that is not bin-centered. The magnitudes of its ordinary DFT (with a pure rectangular window) as well as its DFT using the mismatched time window are shown below in Fig. 3.

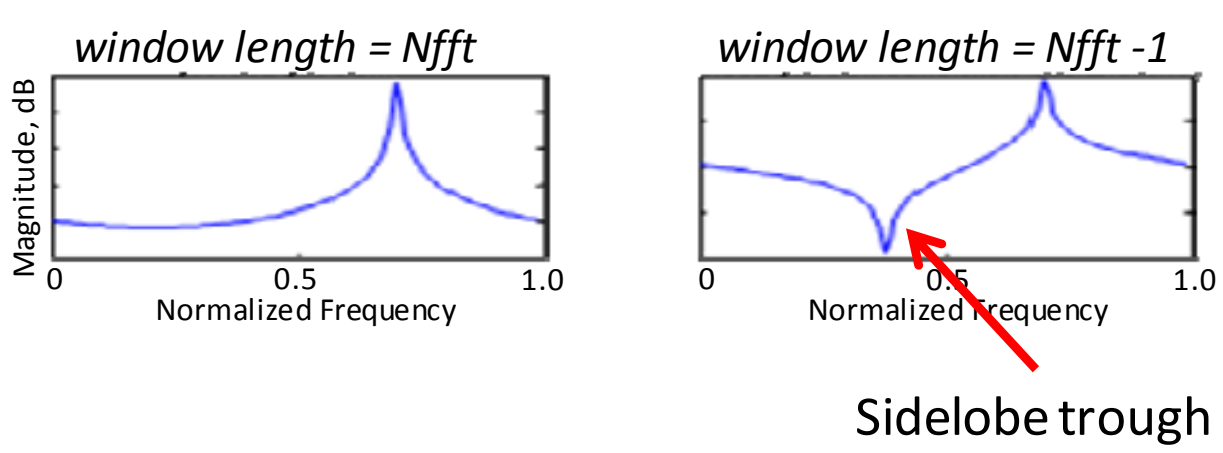

Fig. 3. Power spectra for bin-offset CW signal using both a rectangular window (left) and a mismatched time window (right).

Since the signal's frequency is not bin-centered, we observe significant spectral leakage. When using the mismatched time window, the spectral energy is redistributed so as to form a null. More interestingly, as portrayed in Fig. 4, the location of this null is related to the degree in which the signal's frequency is bin-offset.

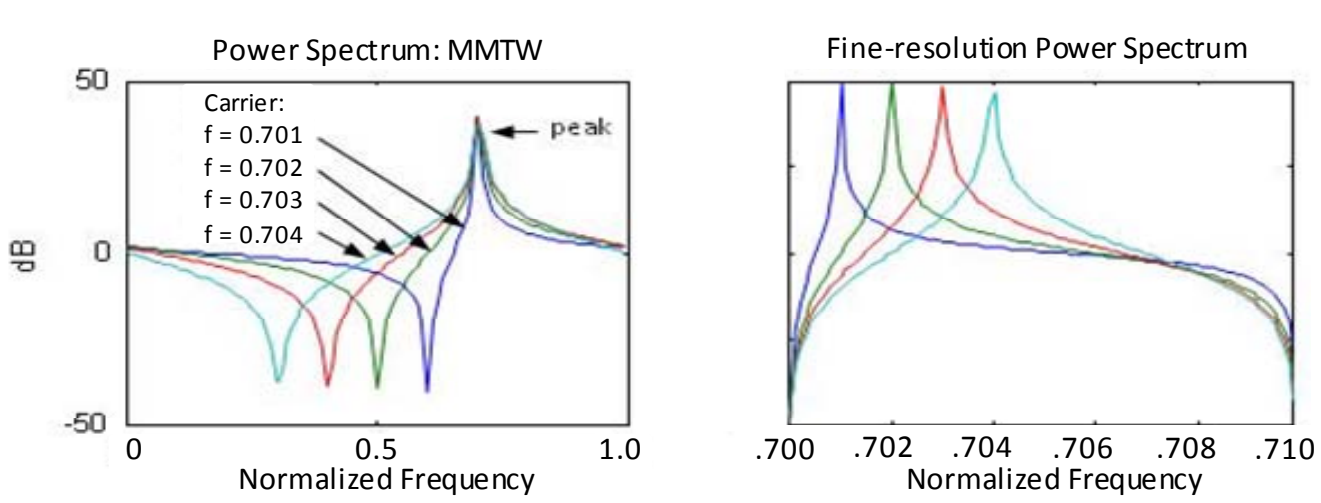

Fig. 4. Power spectra for four signals with slightly different bin-offset frequencies.

The plot on the left shows the DFT of four signals (at normalized frequencies 0.701, 0.702, 0.703, and 0.704). In this case, the bin-center is at 0.700 and the bin-width is 0.010. Now observe that the distances between the peak and each null (moving from right to left, wrapping around if necessary) are proportional to the bin-offset frequencies. More precisely, the following relationship holds:

$$\frac{\text{peak} - \text{null}}{\text{IQ sample rate}} = \frac{\text{bin offset}}{\text{bin width}} \tag{3}$$

A careful mathematical derivation of this proportionality is presented in Section IV.

In the case of Fig. 4, we can calculate the bin-offset of the red signal (0.703) as follow:

$$\begin{aligned} \text{bin offset} &= \text{bin width} \times \frac{\text{peak} - \text{null}}{\text{IQ sample rate}} \\ &= (0.01) \times \frac{0.7 - 0.4}{1} \\ &= 0.003 \end{aligned} \tag{4}$$

The plot on the right side of Fig. 4 shows the inverse magnitudes of the four signals over a fine frequency scale that is derived from (3). Another example of the dual coarse-scale/fine-scale representation of a bin-offset signal is shown in Fig. 5.

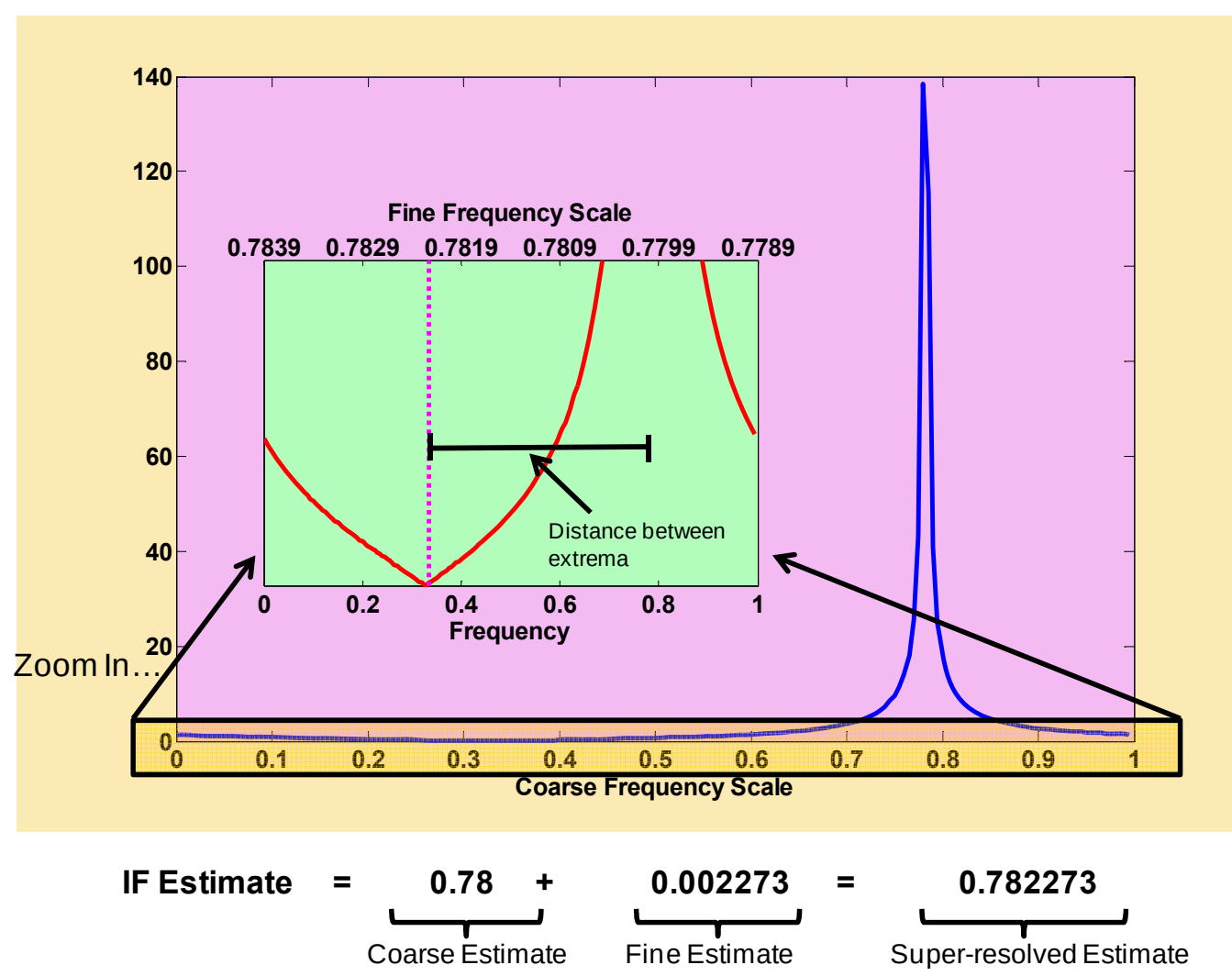

Fig. 5. Coarse and fine frequency scale representations of the DFT of a signal using a mismatched time window.

In this example, the maximum of the spectrum occurs at the bin corresponding to normalized frequency 0.78. By examining the location of the null (shown in the green inset) and applying (3), we calculate the bin offset as 0.002273. Adding this offset to the coarse estimate yields the super-resolved frequency estimate of 0.782273.

Since each column of a spectrogram is precisely equal to a DFT, using a time mismatched window will allow for the extraction of fine frequency modulation from narrowband communications signals. Given a spectrogram or spectrum of a wide-band environment, a user can identify a signal of interest, band-pass filter it, and then apply an MMTW spectrogram in order to super-resolve the underlying modulation. A schematic depiction of this process is shown below in Fig. 6. In the next section, we demonstrate the MMTW on various simulated and laboratory signals.

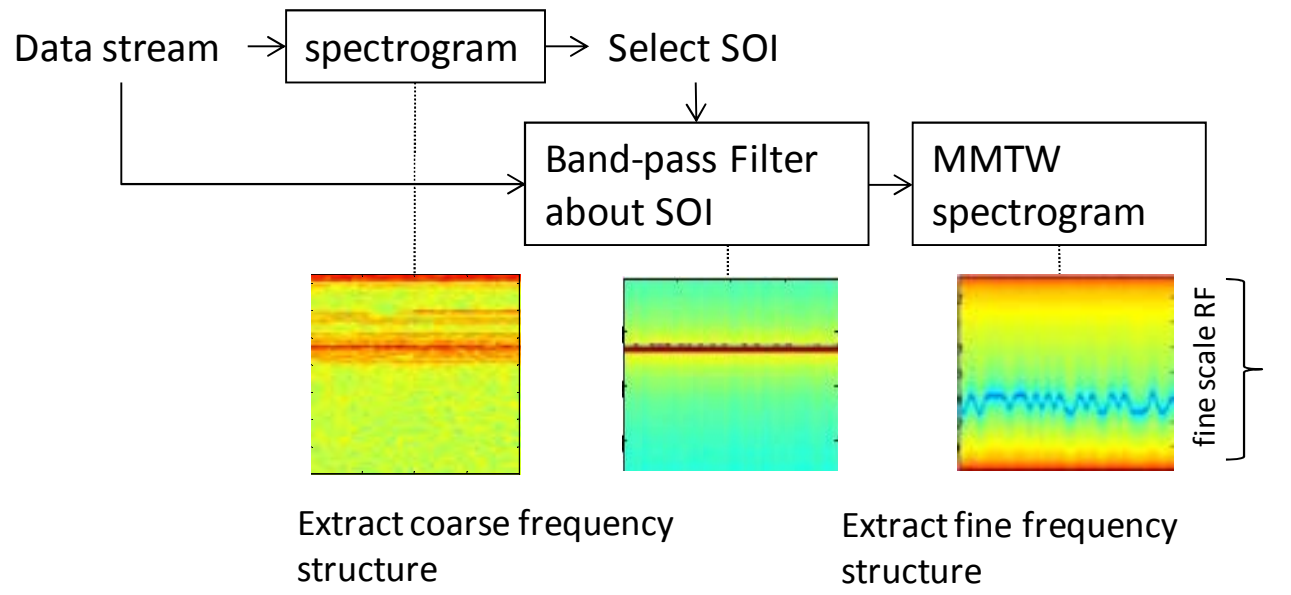

Fig. 6. Flowchart of MMTW Spectrogram Processing

## III. EXPERIMENTS

Our first experiment consists of a simulated FSK signal in a wide-band environment with normalized bandwidth 1. The modulation bandwidth is $1 \times 10^{-4}$, which is considerably smaller than the DFT bin-width of $1 / 512 \approx 2 \times 10^{-3}$. Nevertheless, the MMTW spectrogram is successful at recovering the FSK sequence as shown below in Fig. 7.

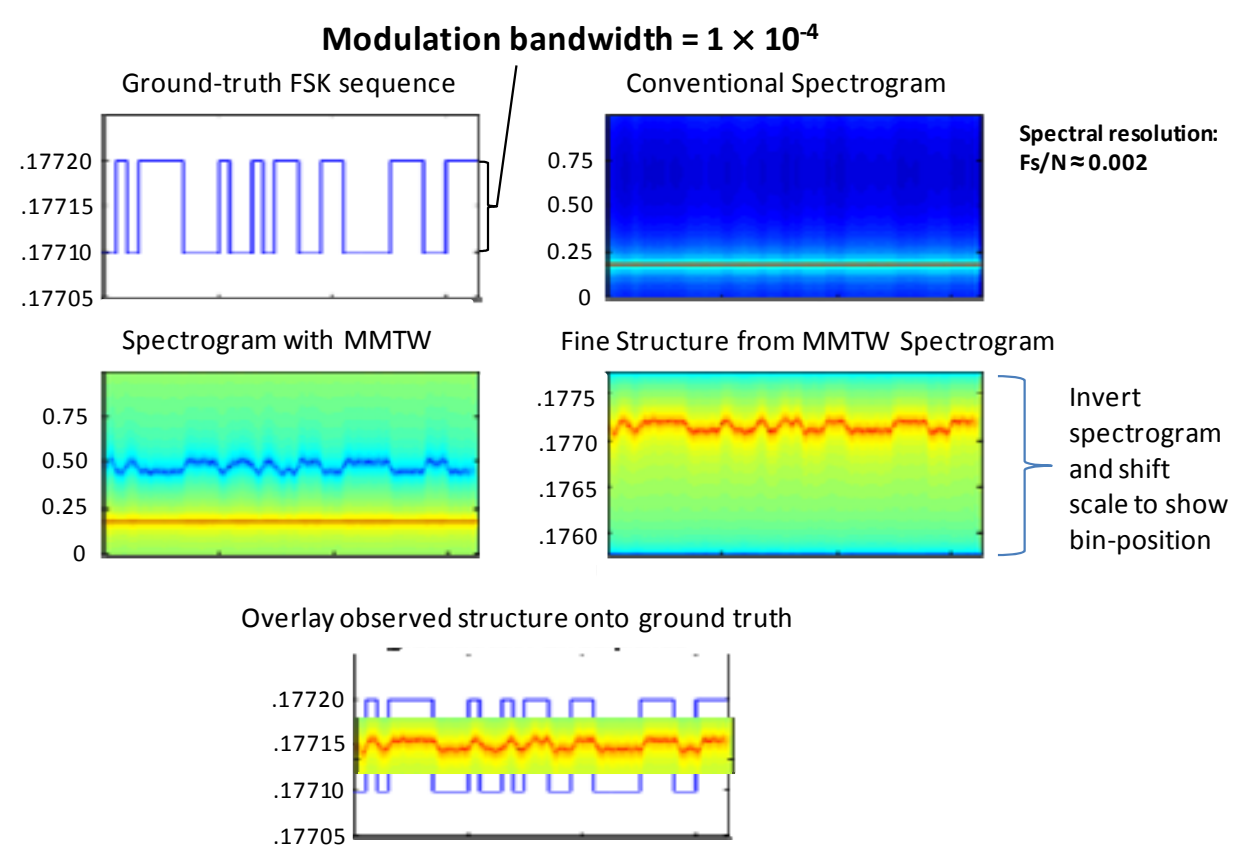

Fig. 7. Extraction of FSK sequence from simulated narrowband signal.

For the next experiment, we examined a signal in a dense environment as shown in the top panel of Fig. 8. After band-pass filtering the signal and applying the MMTW spectrogram, we extracted the underlying modulation shown in the middle panel. For comparison, the modulation structre was also extracted via conventional means, with the band passed instantaneous frequency. The two approaches produced consistent results.

A third shows the effectiveness of the MMTW spectrogram with respect to direction finding. Given an antenna(s) that is mechanically (or electronically) traversing a circular path, it is possible to calculate a signal's direction of arrival (DOA) using Doppler shift [4]. The direction of arrival is simply the phase shift of the Doppler induced sinusoidal modulation. Unfortunately, due to the extremely fast speed of light, it is impossible to detect this modulation on a traditional spectrogram. However, by using the MMTW spectrogram, it is possible to determine DOA from mere visual inspection. In Fig. 9, we show a simulated signal that is received by a Doppler antenna as well as the super-resolved estimate of the Doppler modulation. By using a very straight-forward algorithm, we were able to estimate DOA to within a few degrees.

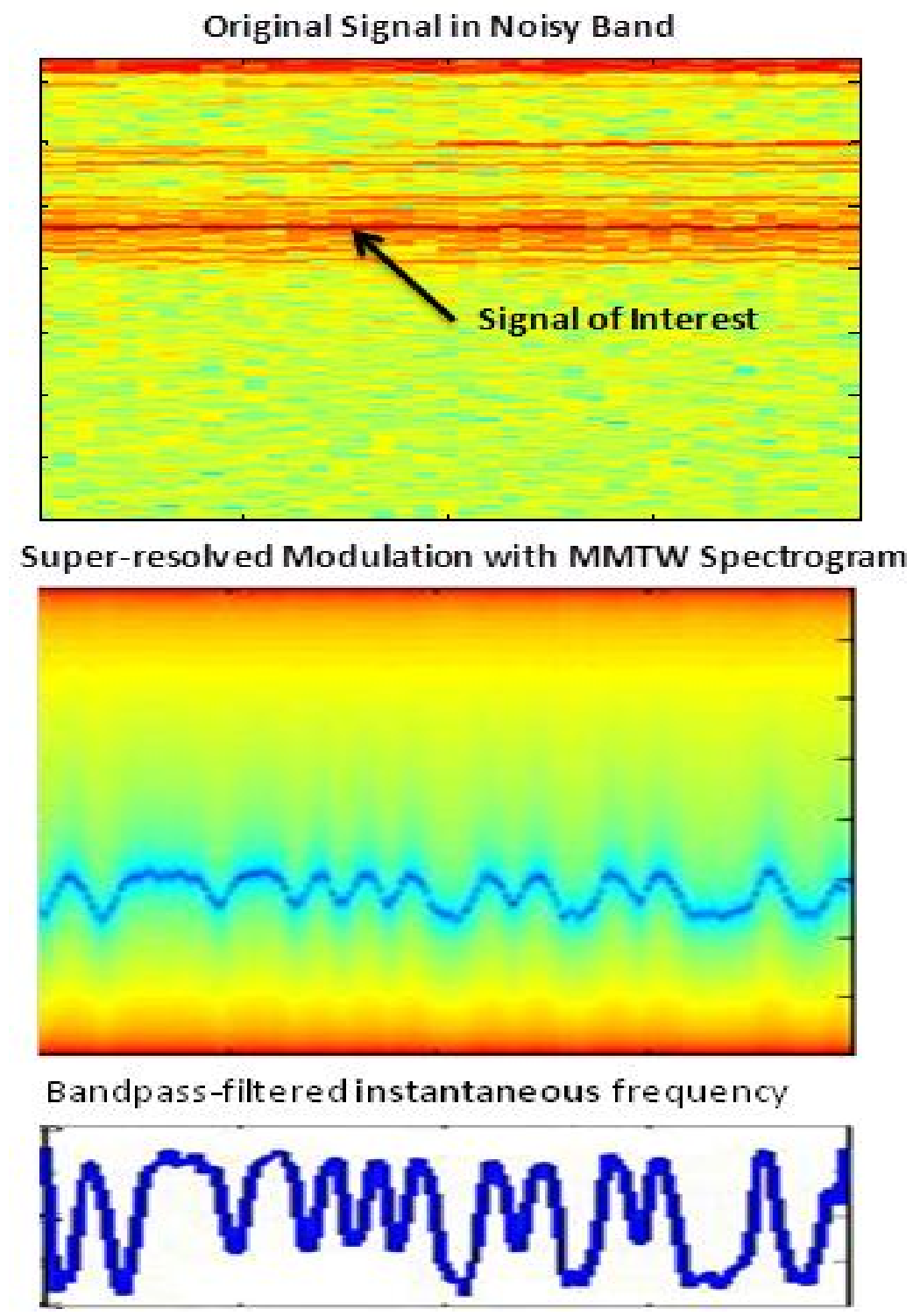

Fig. 8. The effectiveness of the MMTW in dense signal environments.

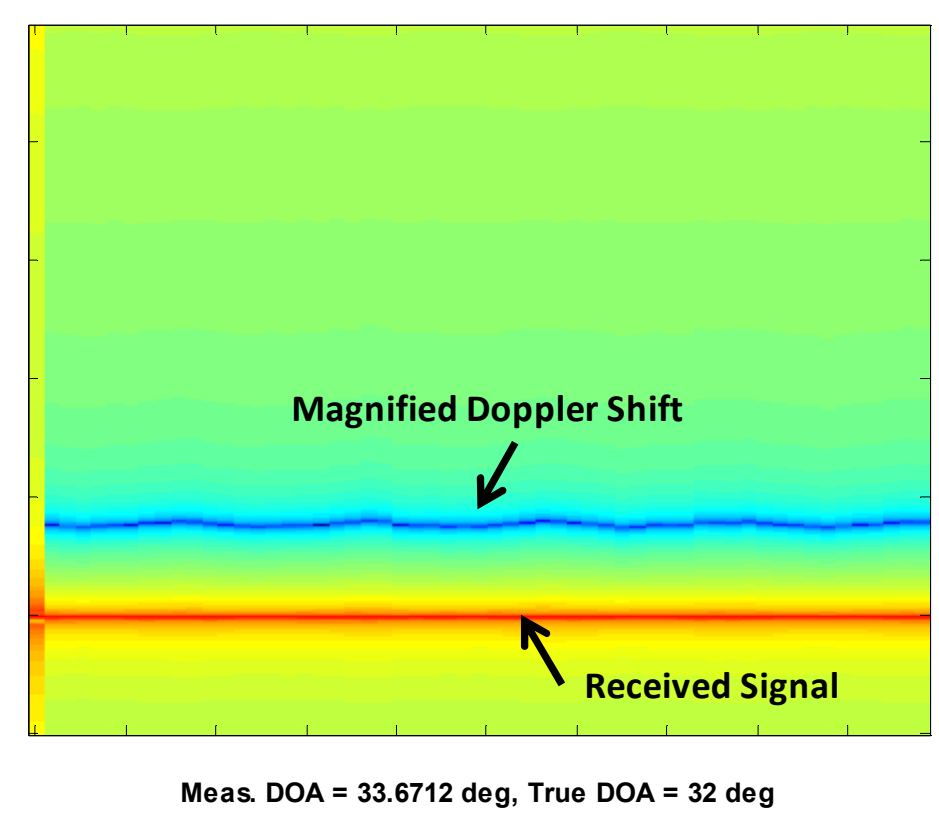

Fig. 9. Estimating DOA by exposing Doppler induced modulation.

Our final experiment involves the use of the MMTW to expose the fine frequency structure of cricket chirps from a sound recording. As is shown in Fig. 10 (a), a raw spectrogram of cricket chirps does little to help us visualize individual chirps. In order to achieve super-resolution, we first up-sampled and interpolated the signal by a factor of 8 and then applied an MMTW spectrogram, which is shown in Fig. 10 (b). Lastly, as shown in Fig. 10 (c), we isolated a single cricket chirp and generated an MMTW spectrogram that not only shows the downward chirp, but some of the frequency structure that is present when the cricket returns its legs to the chirp starting position. Such spectral structure may be invaluable to a biologist studying cricket anatomy.

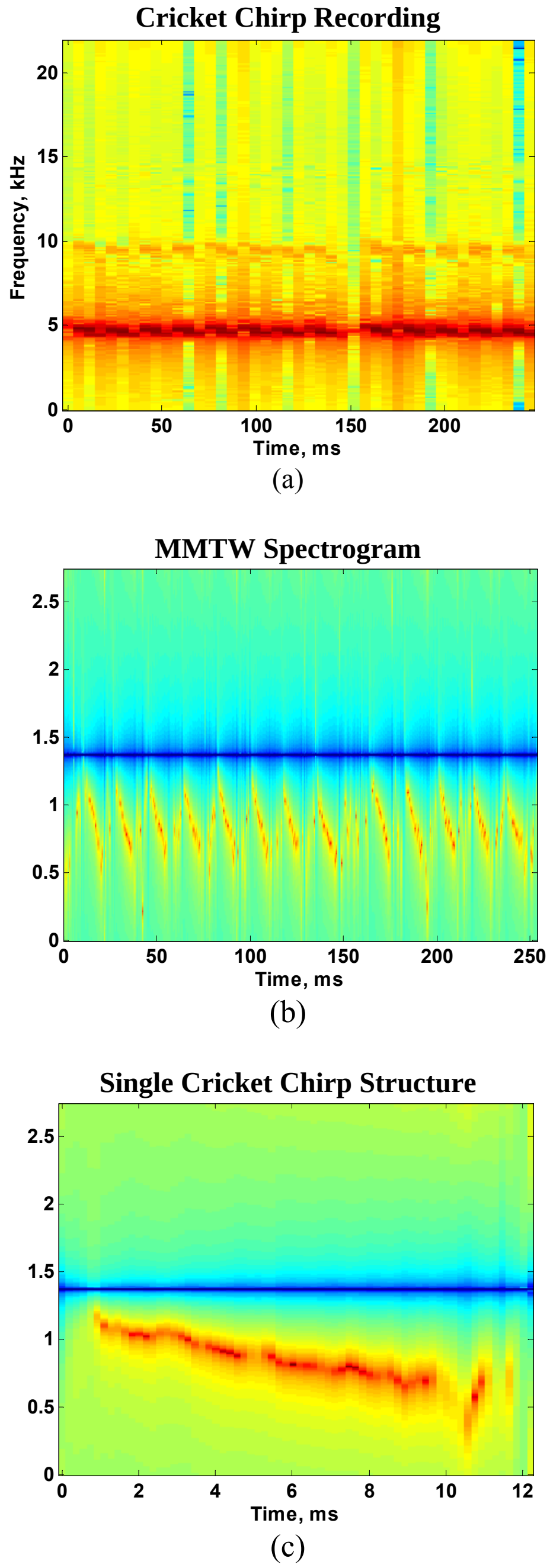

Fig. 10. (a) Traditional spectrogram of cricket chirp sampled at 44.1 kHz. (b) MMTW spectrogram of several cricket chirps. (c) MMTW spectrogram of a single cricket chirp. The frequency scales in the bottom two panes show the amount of bin offset in kHz and do not represent absolute frequencies.

## IV. Mathematical Derivation

To derive the technique presented in this paper, assume $x[n] = A\exp(j2\pi(k_0 + \alpha)n/N)$ is a single tone signal over $N$ samples with complex amplitude $A$, and a non-bin-centered frequency $k_0 + \alpha$ where $k_0 \in \{0,\cdots, N-1\}$ and $\alpha \in [0,1)$ is the bin offset. Our objective is to estimate both $k_0$ and $\alpha$. In order to do this, we first calculate the discrete Fourier transform of the above non-bin-centered signal, which is:

$$X(k) = \sum_{n=0}^{N-1} A\exp\left(j2\pi(k_0+\alpha)n/N\right)\exp(-j2\pi kn/N) = A\frac{\exp\left(j2\pi(k_0+\alpha-k)\right)-1}{\exp\left(j2\pi(k_0+\alpha-k)/N\right)-1}$$

The second equality comes from the fact that the Fourier sum happens to be a geometric series. If we subtract the value of $x[0]$ (i.e., apply the mismatched time window) from the above spectrum and compute its magnitude, we obtain:

$$|X(k)-x[0]| = \left|\frac{A\left(\exp\left(j2\pi(k_0+\alpha-k)\right)-1\right)}{\exp\left(j2\pi(k_0+\alpha-k)/N\right)-1} - A\right|$$

$$= \left|\frac{A\left(\exp\left(j2\pi(k_0+\alpha-k)\right)-\exp\left(\frac{j2\pi(k_0+\alpha-k)}{N}\right)\right)}{\exp\left(j2\pi(k_0+\alpha-k)/N\right)-1}\right|$$

$$= \left|\frac{A\exp\left(j\pi(k_0+\alpha+k)\left(1+\frac{1}{N}\right)\right)}{\exp\left(j2\pi(k_0+\alpha-k)/N\right)-1}\sin\left(\pi(k_0+\alpha-k)\left(1-\frac{1}{N}\right)\right)\right|$$

We note that the above magnitude is equal to zero if, and only if, the sine term is zero, which will happen if, and only if, the quantity $(k_0 + \alpha - k)(1 - 1/N) = (k_0 + \alpha - k)(N - 1)/N$ is equal to an integer. As a consequence, the maximum value of $Y(k) = 1 / |X(k) - x[0]|$ will occur at the integer value of $k$ where the fractional part of $(k_0 + \alpha - k)(1 - 1/N)$ is closest to zero. To find the value of $k$, let's assume for simplicity that the value of the bin-offset $\alpha$ falls on a uniform discrete grid with spacing $1/(N-1)$, i.e. assume $\alpha = r / (N - 1)$ for some integer $r$. Further suppose that there exists some integer $m$ such that

$$\left(k_0+\frac{r}{N-1}-k\right)\left(\frac{N-1}{N}\right) = \frac{(k_0-k)(N-1)+r}{N} = m \quad (5)$$

We seek to solve the above equation for $k$. Rearranging terms gives us that:

$$k_0 - k = \frac{Nm-r}{N-1} = \frac{(N-1)m+(m-r)}{N-1} = m + \frac{m-r}{N-1} \quad (6)$$

Since $k_0 - k$ and $m$ are integers, then so must be the quotient $(m - r)/(N - 1)$. This implies that $m \equiv r \pmod{N-1}$. Solving the above for $k$ yields:

$$k = k_0 - m - \frac{m-r}{N-1} \quad (7)$$

Based on the modular relationship we just deduced, we know that $m = r + q(N - 1)$ for some integer $q$. Substituting this into (7) gives:

$$k = k_0 - r - q(N-1) - \frac{r+q(N-1)-r}{N-1} = k_0 - r - qN$$

In other words, $k \equiv k_0 - r \pmod{N}$. Recalling that $k_0$ is the location of the spectral peak, $k$ is the location of the null, and $r$ is the amount of bin-offset modulo $N - 1$, we see that this relation is precisely equivalent to (3).

For a general $\alpha \in [0,1)$, we find that the minimum value of $|X(k) - x[0]|$ will occur approximately $\alpha N$ bins away from the bin where the maximum of the digital spectrum occurs. This suggests that the location of this minimum can be used to estimate the true frequency of the tone $x[n]$ to within a factor of $O(1/N^2)$. In the following section, we use the Cramér-Rao Bound for frequency estimators to show that this is asymptotically the best super-resolution that we can hope to achieve.

## V. Relationship to the Cramér-Rao Bound

It is not difficult to show (see [5]) that any unbiased frequency estimator that is based on $N$ uniformly spaced IQ samples of a complex sinusoid satisfies the following Cramér-Rao Bound:

$$\begin{aligned} \operatorname{var}(\hat{f}) &\geq \frac{3\sigma^2 f_S^{\,2}}{4\pi^2 A^2 N(N+1)(2N+1)} \\ &= \Theta\left(\frac{1}{N^3}\right) \end{aligned} \tag{8}$$

where $A$ is the signal amplitude, $\sigma$ is the noise standard deviation, and $f_S$ is the IQ sample rate .

Superficially, this seems to contradict the high SNR asymptotic error bound we calculated in the previous section, which implies that the error variance is $O(1/N^4)$. This is another example of the paradoxical behavior of the mismatched time window spectrogram. A quick glance at the problem seems to indicate that the frequency estimates obtained by our $N$ samples seem to violate the Cramér-Rao bound.

Alas, there is no magic being performed here. The increased accuracy stems from the fact that our methodology involves pre-filtering a signal in either the analog domain or in the full digital domain (not just one window). Either way, since we're assuming the target signal modulation has bandwidth less than a single DFT bin-width, it follows that the initial filtering yields a linear processing gain over the SNR $A/\sigma$ of

$$\begin{aligned} \left(\frac{A}{\sigma}\right)_{\text{filt}} &= \sqrt{\frac{f_S}{\text{bin width}}}\left(\frac{A}{\sigma}\right)_{\text{orig}} \\ &= \sqrt{\frac{f_S}{f_S/(CN)}}\left(\frac{A}{\sigma}\right)_{\text{orig}} \\ &= \sqrt{CN}\left(\frac{A}{\sigma}\right)_{\text{orig}} \end{aligned} \tag{9}$$

where the constant $C$ is typically on the order of 0.5 to 1. Factoring the processing gain (9) into the CRB (8) yields:

$$\begin{aligned} \operatorname{var}(\hat{f}) &\geq \frac{3 f_S^{\,3}}{4\pi^2 C N^2 (N+1)(2N+1)}\left(\frac{\sigma}{A}\right)^2_{\text{orig}} \\ &= \Theta\left(\frac{1}{N^4}\right) \end{aligned} \tag{10}$$

We can now conclude that our enhanced frequency estimator has a variance that is asymptotically equivalent to the Cramér-Rao bound for high SNR and no further orders of super-resolution are possible.

## VI. Conclusion

In this paper, we have proposed the MMTW spectrogram as a novel strategy for exploiting spectral leakage in order to recover fine frequency structure present in a signal. In additional to presenting a wide variety of interesting examples, we have exhibited a full mathematical justification of the methodology along with a Cramér-Rao bound analysis showing that it asymptotically achieves maximal frequency resolution. While admittedly not the fastest way to super-resolve frequency information, the MMTW spectrogram does offer enhanced spectral visualization and yields incredible theoretical insight into the importance of signal spectral tails.

## References

[1] B. Boashash, "Estimate and Interpreting the Instantaneous Frequency of a Signal – Part 2: Algorithms and Applications," *Proc. of the IEEE*, vol. 80, no. 4, Apr 1992.

[2] J. Hansen, "Selected Approached to Estimation of Signal Phase," Technical Report, University of Rhode Island, 2003.

[3] E. Jacobsen, P. Kootsookos, "Fast, Accurate Frequency Estimators," *IEEE Signal Proc. Magazine DSP Tips & Tricks*, pp. 123-125, May 2007.

[4] D. Adamy, *EW 101: A First Course in Electronic Warfare*. Boston, MA: Artech House, 2001.

[5] D. C. Rife, R. R. Boorstyn, "Single-Tone Parameter Estimation from Discrete-Time Observations," *IEEE Trans. Info Theory*, vol. IT-20, pp. 591-598, 1974.